\documentclass[sigconf,nonacm]{acmart}
\AtBeginDocument{%
  }

\acmISBN{978-1-4503-XXXX-X/2018/06}

\usepackage{multirow}

\begin{document}

\title{Agentic Path Maps: Exposing an Agent’s Traversal Through the Garden of Forking Paths}


\author{Madeleine Grunde-McLaughlin}
\email{mgrunde@cs.washington.edu}
\orcid{0000-0001-7290-068X}
\affiliation{
  \institution{University of Washington}
  \streetaddress{3800 E Stevens Way NE}
  \city{Seattle}
  \state{Washington}
  \country{USA}
  \postcode{98195}
}

\author{Hafsah Usman}
\email{husman@uw.edu}
\orcid{}
\affiliation{
  \institution{University of Washington}
  \streetaddress{3800 E Stevens Way NE}
  \city{Seattle}
  \state{Washington}
  \country{USA}
  \postcode{98195}
}

\author{Daniel Mitroshkov}
\email{dmkov15@cs.washington.edu}
\orcid{}
\affiliation{
  \institution{University of Washington}
  \streetaddress{3800 E Stevens Way NE}
  \city{Seattle}
  \state{Washington}
  \country{USA}
  \postcode{98195}
}

\author{Emily Reif}
\email{emreif@cs.washington.edu}
\orcid{0000-0003-3572-6234}
\affiliation{
  \institution{University of Washington}
  \streetaddress{3800 E Stevens Way NE}
  \city{Seattle}
  \state{Washington}
  \country{USA}
  \postcode{98195}
}

\author{Nino Migineishvili}
\email{ninom@cs.washington.edu}
\orcid{0009-0009-0587-2100}
\affiliation{
  \institution{University of Washington}
  \streetaddress{3800 E Stevens Way NE}
  \city{Seattle}
  \state{Washington}
  \country{USA}
  \postcode{98195}
}

\author{Lakshya Chauhan}
\email{lakshy16@cs.washington.edu}
\orcid{}
\affiliation{
  \institution{University of Washington}
  \streetaddress{3800 E Stevens Way NE}
  \city{Seattle}
  \state{Washington}
  \country{USA}
  \postcode{98195}
}

\author{Kevin Yu}
\email{bigkyu@cs.washington.edu}
\orcid{}
\affiliation{
  \institution{University of Washington}
  \streetaddress{3800 E Stevens Way NE}
  \city{Seattle}
  \state{Washington}
  \country{USA}
  \postcode{98195}
}

\author{Jeffrey Heer}
\email{jheer@cs.washington.edu}
\orcid{0000-0002-6175-1655}
\affiliation{
  \institution{University of Washington}
  \streetaddress{3800 E Stevens Way NE}
  \city{Seattle}
  \state{Washington}
  \country{USA}
  \postcode{98195}
}

\author{Daniel S. Weld}
\email{danw@allenai.org}
\orcid{0000-0002-3255-0109}
\affiliation{
  \institution{Allen Institute for AI}
  \streetaddress{2157 N Northlake Way #110}
  \city{Seattle}
  \state{Washington}
  \country{USA}
  \postcode{98103}
}
\affiliation{
  \institution{University of Washington}
  \streetaddress{3800 E Stevens Way NE}
  \city{Seattle}
  \state{Washington}
  \country{USA}
  \postcode{98195}
}








\renewcommand{\shortauthors}{Anon.}

\begin{abstract}
When an AI agent executes an underspecified task, it embeds choices that may contradict unarticulated user context and preferences. 
However, interrupting the user for clarification on all such decisions is untenable. 
How might agents communicate impactful, but unexplored, paths to the user after execution? 
To investigate, we first distill a design space for communicating alternative paths, supporting the generation of new systems and structuring dimensions of future study.
We then measure the impact of one vital design space dimension on user comprehension: the number of alternatives displayed.  
Through a user study (n=48) in the data analysis domain, we find that participants rarely consider alternative paths of their own volition, but AI overreliance reduces when alternative paths are visualized for them. 
By revealing paths the agent did \textit{not} take, we demonstrate how users can better interpret results and incorporate personal context.

\end{abstract}


\begin{teaserfigure}
  \includegraphics[width=\textwidth]{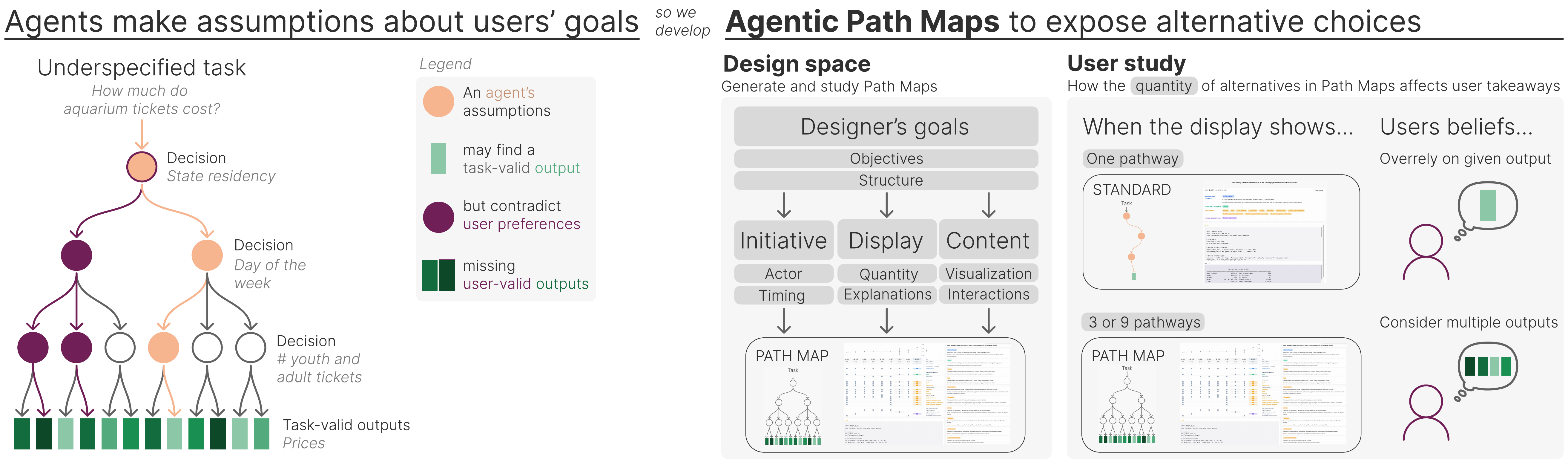}
  \caption{Agentic workflows make assumptions about user's goals to complete underspecified tasks. Although the agent's found output may be valid under the task constraints, it may not be valid given the users' context and preferences. In this paper, we contribute Agentic Path Maps to communicate untaken alternative paths to users. We first distill a space of designs for generating and studying Agentic Path Maps. We then manipulate one dimension, the quantity of paths shown, in an empirical user study. We find that participants overrelied less and were more cognitively engaged when shown alternative paths.}
  \Description{
    Overview diagram linking an agent's decision tree, the Path Map design space, and the user study. On the left is a ticket-purchasing task which branches through decisions about state residency, day of the week, and youth versus adult tickets, ending in prices. A highlighted path marks the agent's assumptions and leads to a task-valid output. The other branches mark choices that contradict user preferences and overlooked outputs valid to the user. In the center are designer goals, showing  objectives and structure underneath. Arrows then point to a Path Map interface paired with a branching diagram and analysis results. On the right are two rows that pair study interfaces with illustrations of users' beliefs. The standard interface shows one pathway and a thought bubble containing one output. The Path Map interface shows three or nine pathways and a thought bubble containing multiple outputs.
  }
  \label{fig:pull}
\end{teaserfigure}

\received{20 February 2007}
\received[revised]{12 March 2009}
\received[accepted]{5 June 2009}

\maketitle

\section{Introduction}
\label{sec:intro}

Consider a teacher using an agent to help plan a field trip to the aquarium, who then finds they owe an unexpected surcharge upon arrival with students. Given that the teacher had taken precautions to carefully verify the agent's found price, how did this error occur? 

When inputting tasks to agents, users specify context that they think might impact the outcomes, but they cannot predict everything. 
The teacher queried "How much do tickets to the aquarium cost for 50 7-9 year olds and 6 adults?", specifying ages because they predicted that youth tickets would be priced lower. 
After the agent returned the price, the teacher reviewed the agent's actions, verified that it was faithful to the website, confirmed that the price was within budget, and requested the agent to purchase the tickets.
However, when arriving at the aquarium, our teacher realized that the agent had incorrectly assumed all students were state residents and purchased tickets accordingly. 
Although the returned price was correct under the task's stated constraints (\textit{task-valid}), our teacher held personal context, knowledge, and preferences that made it incorrect for their particular needs (\textit{user-valid}).

Underspecified and ambiguous tasks, in which a computing system does not know the user-valid subset of task-valid options, are well-recognized as a critical issue both for AI agents~\cite{deshpande2025trail,bansal2024challenges} and other computing fields that interpret user context~\cite{mankoff2007interaction,hu2011characterizing,keyvan2022how,dey2005designing}. 
Gelman and Lopez~\cite{gelman2013garden} coined the term ``garden of forking paths'' regarding specification ambiguity in hypothesis testing, referring to the branching space of possible analysis constructions. 
Users give agents complex tasks in open-world environments, so it is challenging to predict the {\em agent's} ``garden of forking paths'' for that task before execution~\cite{bansal2024challenges}. 
This issue persists even as agents improve: a seemingly ``perfect'' agent that invariably finds a task-valid output still may not meet the user's needs.

For a user to know if a given outcome contradicts their personal context, they must first understand the space of decision points and how alternative choices at those points lead to different outcomes.
If the teacher learned how residency impacts prices when receiving the estimate, they could consider the price that matches their needs before purchasing tickets and without being interrupted as the agent finds prices. 
This simple scenario illustrates a larger key human-AI interaction challenge: \emph{How can we support users in (1) understanding a diverging space of task-valid options for (2) underspecified tasks, while (3) still enabling the agent to execute automatically?} {\em Human understanding} is a key aspect of auditing AI outputs, {\em underspecified tasks} are ubiquitous as no human can anticipate all possible choices~\cite{McCarthy1987SOMEPP}, and {\em automatic agentic execution} mitigates the burden of constant monitoring or responding to excessive interruptions. 

Many methods support some of these goals, but the means of supporting all three remains underexplored. 
In standard practice, an agent makes assumptions during {\em automatic execution}, allowing it to complete {\em underspecified tasks}. An interface may then transparently provide the trace of agent actions for users to review~\cite{pang2025interactive,kazemitabaar2024improving,grunde2026overseeing,chang2026criticality}. 
However, this approach does not explicitly support users in  {\em exploring the space of possibilities}, instead relying on users to notice and verify decisions that they did not think to pre-specify.
Interrupting the user to clarify {\em underspecified options} usefully nudges users to consider {\em diverging options}~\cite{zhang2025clarify,dong2026value,liu2026perspectra}. 
However, interruptions do not enable {\em automatic execution} and can put an undue burden on the user if many assumptions must be clarified in complex tasks.
By modeling the user, personalization aims to reduce the need for {\em users' verification of assumptions} by inferring user preferences when {\em tasks are underspecified}, enabling {\em automatic task completion}~\cite{lam2026just,li2026survey,zhuang2024hydra}. 
However, there will inevitably be some context of which the user model is unaware~\cite{schmit2018human}, so modeling cannot eliminate the need for users' {\em interpretation of alternatives}.

In this paper, we instead consider agentic systems that explicitly represent multiple action paths and present them to users for post-hoc review. We propose \textbf{Agentic Path Maps}: displays that enable users to review (1) alternative pathways with task-valid options, to narrow an (2) underspecified decision space down to their own user-valid subset of outcomes (3) after the agent executes. 
Towards this end, we contribute a generative design space for Agentic Path Maps and empirical results demonstrating how viewing alternative paths improves human sensemaking of agent actions.

With a large quantity of decision points and alternative paths, the information to communicate can be dense and complex, making it non-trivial to effectively communicate alternative paths to people. 
Rather than starting from scratch, we contribute a generative design space by gathering inspiration from other domains that display multiple options, such as counterfactual explanations, multiverse analysis, and decision trees. 
We reviewed 2,402 unique papers retrieved by ten keywords from related domains and filtered them to 56 papers with mixed-initiative artifacts displaying alternatives.

From these papers, we distill a design space of eight dimensions: \textit{objectives for communicating multiple items}, \textit{the structure of the multiple items}, \textit{if users or agents take initiative}, \textit{initiative timing}, \textit{the quantity of content to display}, \textit{additional explanatory content}, \textit{the visualization}, and \textit{user interactions}. 
As the design space was informed by multiple domains, we highlight the elements for each dimension specifically relevant to Agentic Path Maps. 
We define connections between the dimensions, such that one can build an Agentic Path Map starting with one's goals, then progressing through the dimensions following relevant connections. 
The design space can also structure experiments manipulating one dimension to study the effect on objectives and define more detailed recommendations.

We next investigate the impact of the \textit{quantity of content displayed} dimension on the \textit{improving understanding} goal on two levels with an empirical study (n=48). 
Our preregistered hypotheses manipulate whether participants see one or multiple paths to confirm whether displaying alternative paths improves user understanding.  
In additional analyses, we manipulate whether participants view nine paths or a sample of three. 
We study a data analysis agent~\cite{ai22025asta}, as data analysis tasks have measurable quantitative outputs and we can build on prior work from multiverse analysis. 
In multiverse analysis~\cite{simonsohn2015specification,steegen2016increasing,gelman2013garden,simmons2011false,schweinsberg2021same}, analysts define a set of defensible analytic choices and then test multiple combinations, sometimes leading to substantially divergent outcomes.

As multiverse analysis can produce results with highly variable, even contradictory, estimates, it can serve as a petri dish to study human interpretation of AI outputs~\cite{gu2024blade}. 
We first elicit participants' prior beliefs about a data analysis question, then display AI-generated analyses, and subsequently elicit their posterior beliefs. 
Although multiverses have no optimal takeaway, we consider it misleading if the participant's posterior swings towards the extreme ends of the output distribution. The baseline condition displays a single analysis producing an extreme output to assess anchoring to one of many task-valid results. 
We measure overreliance as the Weight of Advice users place on extreme outputs~\cite{ibrahim2025measuring,passi2025addressing,bailey2023meta}. We additionally measure Intrinsic, Extraneous, and Germane Cognitive Load~\cite{krieglstein2023development}.

Participants overrelied less and ended with posterior beliefs closer to the multiverse median when shown multiple paths. 
When asked what they would attempt next, participants rarely suggested specific alternative analysis choices of their own volition, instead anchoring to the AI's displayed decisions. 
Although they report increased Intrinsic and Extraneous Cognitive Load when viewing multiple alternatives, we also find a small effect of increased Germane Cognitive Load, a measure of beneficial cognitive engagement.
When displaying nine results rather than a sample of three, most benefits and drawbacks diminished, except for Germane Cognitive Load. We find instead a small effect that participants reported higher Germane Cognitive Load when shown the sample, indicating that displaying a sample of pathways is a promising middle ground. 
Our results demonstrate empirical evidence that providing users with a task-valid output can still be misleading, as users anchor to the agent's chosen path. 

We therefore encourage agentic system designers to consider if interruptions are necessary to adapt to underspecified tasks, or if a suite of possibilities can instead be displayed post-hoc.
We discuss the conditions under which we see Agentic Path Maps as most useful, how studying them can redesign evaluation approaches for underspecified tasks and uncertain human overseers, and promising regions for expanding the design space. 
Thus, we contribute: 
\begin{itemize}
    \item A proposal for the development of Agentic Path Maps to support post-hoc communication of alternative paths.
    \item A design space for communicating alternative paths to human reviewers that is generative for interfaces and informative for structuring future studies.
    \item Empirical results that human reviewers overrelied on task-valid outputs when they did not view alternative paths. 
    \item Empirical results that increasing the number of alternatives shown (i.e., the quantity dimension of our design space) has diminishing returns. 
\end{itemize}

\section{Related work}
\label{sec:rw}

We situate our work among methods for addressing task ambiguity (\ref{sec:rw-ambiguity}), multiverse analysis (\ref{sec:rw-multiverse}), and interpretability (\ref{sec:rw-interpretability}).

\subsection{Methods for addressing ambiguity}
\label{sec:rw-ambiguity} 

Agents must understand a goal, plan its completion, then implement the plan's actions~\cite{russell1995artificial}, but an ambiguous goal makes the completion criteria unknown. 
Ambiguous goals are a concern for both modern agents~\cite{deshpande2025trail,bansal2024challenges,pang2025interactive,chang2026criticality} and other user-intent based applications~\cite{mankoff2007interaction,hu2011characterizing,keyvan2022how,dey2005designing}. In fact, ``considering uncertainty about a user's goals'' is a critical factor in Horvitz's Principles of Mixed-Initiative User Interfaces~\cite{horvitz1999principles}. 
Many approaches aim to address ambiguity for agents. 

Techniques like Tree of Thought~\cite{yao2023tree} and Graph of Thoughts~\cite{besta2024graph} apply when an ambiguous path leads to one correct answer, while decision-theoretic approaches predefine one or more qualities to optimize in the ambiguous space~\cite{einhorn1986decision,agarwal2025autodiscovery,dai2010decision}.
For instance, AutoDiscovery adapts to underspecified scientific queries by optimizing for Bayesian surprise as a proxy for novelty~\cite{agarwal2025autodiscovery}. 
However, these approaches cannot address cases where ambiguity rests in a user's context and preferences. 

Personalization approaches analyze user behavior over time to model the user's intents~\cite{lam2026just,li2026survey,zhuang2024hydra,brown1998utility}. 
For instance, the tool-using Poppins system implements Just in Time Objectives by passively observing the user over time, interpreting their specific objectives, and inputting that into the model~\cite{lam2026just}.
This approach can smooth interactions and provide more targeted results. However, models have limited knowledge of the user, may induce privacy concerns, and may not work if context abruptly changes. 

Other approaches, from which we drew the introductory example, display assumptions for users to correct~\cite{pang2025interactive,kazemitabaar2024improving,grunde2026overseeing,chang2026criticality}. 
For instance, the CRITICALITY system restructured the trace of agent actions to surface assumptions in retrieving information from large documents explicitly~\cite{chang2026criticality}. 
Nevertheless, the burden remains on the user to not only notice the contradiction, but also recognize what alternatives are available.

Systems may also interrupt the user to clarify ambiguity~\cite{zhang2025clarify,zhang2024ask,mozannar2025magentic,boutilier2002pomdp,schlangen2004causes}, or support constant user involvement with an interactive interface~\cite{huang2025narrative,angert2023spellburst,riche2025instruments}.
This interaction type works well when the user is amenable to interruption or constant oversight, but many users do not want to sit and watch an agent as it executes~\cite{grunde2026overseeing}. Much work investigates the optimum interruption rate to avoid excessive burden~\cite{dong2026value,chan2017semantically,kraus2020effects}, but this approach inherently avoids some interruptions. 

Most similar to us are systems for navigating multiple outputs from an AI, such as Luminate for creative writing~\cite{z1k} and Sensecape for sensemaking~\cite{suh2023sensecape}.
Concurrent work by Ye et al. conceptualizes Large Language Model outputs as multiverses for tasks like philosophical queries and poetry writing~\cite{ye2026navigating}. 
Our design space incorporates these designs and more to distill dimensions for displaying alternative agent actions in an open-world environment.

\subsection{Multiverse Analysis}
\label{sec:rw-multiverse}

We draw from prior work in multiverse analysis to empirically study the effect of showing alternative paths. 
When testing a hypothesis, researchers reckon with decisions such as defining outliers, covariates, and statistical models~\cite{wicherts2016degrees}. Following different routes along the ``garden of forking paths''~\cite{gelman2013garden} can lead to substantially different outcomes~\cite{hall2022survey,gelman2013garden,simmons2011false,schweinsberg2021same}, so multiverse analysis tests a set of defensible combinations of analytic choices to review the robustness of an effect~\cite{simonsohn2015specification,steegen2016increasing}. 
Bespoke systems, like the Boba~\cite{liu2021boba} and Milliways~\cite{sarma2024milliways} dashboards, support the creation and interpretation of multiverse analyses~\cite{sarma2023multiverse,young2017model,dragicevic2019increasing}.
Other studies provide planning assistance while users create a data analysis~\cite{gu2024data,raghunandan2024lodestar}, and the BLADE benchmark evaluates AI on a branching decision space of possible analyses~\cite{gu2024blade}. 
 
More recent work uses AI to build multiverses~\cite{bertran2026many,miao2026agentic}, but there remain critical challenges.  
Poorly justified or a high quantity of alternatives can overwhelm and hide meaningful effects, and human interpretation remains essential as there is no "correct" multiverse or outcome~\cite{del2021traveler,sarma2024milliways}. Furthermore,
AI-created multiverses could facilitate $p$-hacking by producing many outcomes from which researchers choose desired results~\cite{head2015extent}.  
Rather than designing an AI-backed system for multiverse creation or improving upon bespoke systems for human-created multiverses, we scaffold our empirical study with an expert-validated decision space from BLADE.

\subsection{Interpretability and Explainable AI}
\label{sec:rw-interpretability}

Improving transparency and appropriate reliance on generative AI is a critical challenge~\cite{shavit2023practices,ibrahim2025measuring,kim2024im,bo2025rely}.
Towards these ends, Explainable AI (XAI) and interpretability methods aim to improve understanding of how models arrived at their answers~\cite{weld2019challenge}. 
XAI must be carefully designed to foster appropriate reliance~\cite{buccinca2021trust,vasconcelos2023explanations} and to address the challenge that Human-AI teams rarely perform better than either humans or AI alone~\cite{bansal2021does,vaccaro2024combinations}. 
Although the faithfulness of agent traces is easier to verify than that of black-box model features~\cite{agarwal2024faithfulness}, challenges arise from the overwhelming quantity of information to parse~\cite{shavit2023practices,feng2025regulatory,bansal2024challenges}.
Agentic Path Maps help users interpret not only the agent's main course of action, but also the broader decision and outcome spaces. 
Our design space draws on counterfactual explanations~\cite{stepin2021survey} and decision tree~\cite{blockeel2023decision} interpretability methods.

\section{Design space methodology}
\label{sec:dsmethods}
We create a design space for systems that present multiple items (\ref{sec:dsmethods-terms})
by collecting papers (\ref{sec:dsmethods-papers}) and defining themes (\ref{sec:dsmethods-coding}).

\subsection{Terms}
\label{sec:dsmethods-terms}
For clarity, we define terms used in the design space. The system is the interface by which users query agents and receive outputs. Multiples are decision points, alternatives, or outcomes for which systems display multiple options. The designer is the person building the system. The user is the person interacting with the system. The agent is an AI agent for the purposes of our design space but can represent any automated process. A path is a set of decisions leading to one outcome.

\subsection{Paper collection}
\label{sec:dsmethods-papers}

We first brainstormed keywords for related domains, supplemented by querying Gemini~\cite{gemini} and Asta~\cite{asta}. Three authors selected the final set of ten keywords: multiverse analysis, counterfactual explanations, divergent thinking, multi-agent debugging, degrees of freedom, planning assistance, user intent modeling, tree-of-thought (Asta), multi-criteria decision-making (Asta), and decision trees (Gemini). For broader coverage, we also appended the words ``user,'' ``human,'' and ``survey'' for a total of forty search terms.
We then retrieved the top 100 papers for each term with the Semantic Scholar API~\cite{kinney2023semantic}. This initial step surfaced 4,000 papers overall and 2,402 unique papers with abstracts.

As many irrelevant papers surfaced (e.g., environmental publications for ``Tree-of-thought''), we used the Claude API (claude-opus-4-5)~\cite{claude-opus-4-5} to further filter for relevance. After hand-labeling the 400 surfaced papers from the keywords ``multiverse analysis'', ``tree of thought,'' and ``divergent thinking'' as ``relevant,'' ``unsure,'' and ``not relevant,'' we prompt-engineered on the ``multiverse analysis'' keyword until there were no discrepancies between those Claude classified as irrelevant and those human coders deemed relevant. We tested our prompt on the other two keywords and found only three errors on edge cases. We then applied the filtering to data from the remaining keywords. All labeled ``relevant'' or ``unsure'' proceeded to the next round of filtering. To confirm no essential paper was missed, the first author read through all ``non-relevant'' titles, finding three additional relevant publications. In total, 1,080 papers moved to the subsequent stage.

Next, the first author reviewed all titles and abstracts to rate papers by relevance on a five-level scale (``Definitely,'' ``A great option,'' ``If needed,'' ``Not relevant,'' and ``Error''), consulting the full paper as necessary. The top two ranked categories contained 75 papers, which we supplemented with two papers from the first author's prior knowledge~\cite{ye2026navigating,z1k}.
Finally, three researchers co-developed criteria to filter papers as they were read in detail. The paper must include 1) a visual artifact that 2) displays multiple items, 3) is non-deterministic, 4) is mixed-initiative, and 5) requires human interpretation of outcomes. Our final set has 56 papers. 

\subsection{Thematic analysis procedure}
\label{sec:dsmethods-coding}

We followed the thematic analysis procedures described by Braun and Clarke's original~\cite{braun2006using} and adapted~\cite{braun2019reflecting} conceptualizations, particularly codebook thematic analysis. Such approaches are reflexive to the researchers' assumptions and interpretations, putting the researcher in an active role throughout the process. We also consulted previously created generative~\cite{head2021augmenting,heer2009design} and descriptive~\cite{bae2022making,lai2023towards,epstein2020mapping} design spaces and considerations in HCI.

Three authors took part in the thematic analysis procedure. First, we created an initial set of categories based on who, what, where, when, why, and how categories, refined through three rounds of paper annotation.  Second, we deductively coded all papers with this protocol, which is available in Supplementary Materials\footnote{\url{https://github.com/madeleinegrunde/agentic_path_maps}}. 
We tagged relevant text with categories and short descriptions. Third, each researcher independently, then together, inductively coded hierarchical themes from the annotations within each code category. Next, we deductively re-categorized all relevant text, then inductively updated the themes to improve coverage. Finally, we repeated the deductive then inductive cycle within each dimension. All coding was completed by hand.

\input{tables/design_space}
\section{Design space}
\label{sec:ds} 

The design space contains four high-level categories: the designer's goals (\ref{sec:ds-goals}), the system's initiatives (\ref{sec:ds-initative}), the content shown to the user (\ref{sec:ds-content}), and the display (\ref{sec:ds-display}). It encapsulates the varied systems from our reference literature, so we indicate the subset relevant to Agentic Path Maps in Table~\ref{tab:design_space}. The design space provides the structure to design our user study and generate its stimulus (\ref{sec:ds-utilizing}).

\subsection{Goals}
\label{sec:ds-goals} 
The designer has objectives (\ref{sec:ds-goals-objectives}) for communicating multiples and knows the structure of the multiples they want to communicate (\ref{sec:ds-goals-structure}).

\subsubsection{Objectives: Reasons for communicating multiples.}
\label{sec:ds-goals-objectives} 
There are several benefits to communicating multiple outputs, rather than just one output. 

\vspace{0.5em}
\noindent\textbf{Building understanding.} Multiples can improve user understanding of the decision space~\cite{zo,z1k,zm,zc,z18}, outcome space~\cite{z1k,z1a,ze}, agent's process~\cite{zo,z1l,z1b,za,z1g,z4}, or underlying reality~\cite{ze,zq,z1i,zi}. Users can also improve their understanding of the robustness of the proposed outcome~\cite{zl,zf,ze,zj,zp,zz} and the sensitivity of decision points on the outcome~\cite{ze,z4,z1k,zo,z13}. Visually displaying multiples can improve understanding of large-scale information~\cite{ze,z9,zi,z1n,zt}.

\vspace{0.5em}
\noindent\textbf{Integrating user perspective.} 
Users may hold knowledge that can verify or improve autonomously created outputs~\cite{zy,zm,zb,ze,zo,za} or decision points~\cite{ze,z1a}. 
Users may also have personal context and preferences they cannot articulate before engaging with multiples~\cite{za,zc,z5,zi}. By reviewing multiple options, users will understand what ``feels right''~\cite{z5,zb,zm} or will introspect on what is important to them~\cite{z1k,z1j}. 
Multiples support this reflection and increase users' control~\cite{zi,zv,z17,zd}. 

\vspace{0.5em}
\noindent\textbf{Improving task performance.} Including multiples can improve task performance metrics~\cite{zc,z1m,zm,zd,z1d,zu}.

\vspace{0.5em}
\noindent\textbf{Fostering creativity.} Including multiples can spur creativity~\cite{z1l,z1k,z10,z1m}, avoid writers block~\cite{z1m,z1l,z19,z12,z5} and increase exposure to additional viewpoints~\cite{zj,z5,z2}. Showing multiples supports divergent thinking to avoid early convergence on sub-optimal outcomes~\cite{z5,zh,z7,z1m,z1k,z17}. 

\vspace{0.5em}
\noindent\textit{Connections.}  
\textit{Creativity}, \textit{task performance}, and \textit{user perspective} are all well supported by \textbf{interactions} for \textit{integrating perspectives}. \textit{Improving understanding} often must support a high quantity of information, which can be supported by the \textit{navigation}, \textit{eliciting detail}, and \textit{aggregation} \textbf{interactions}. 
\textbf{Explanations} are most used when supporting the goal of \textit{improved understanding}, such as by displaying a \textit{sample} and \textit{trend aggregation} for a large scale of information.
\textit{Hierarchical and independent} \textbf{structures} often integrate \textit{user perspective} and \textit{improving understanding}.
\textit{Outcome sets} are often used for \textit{improve creativity} and incorporate \textit{user perspectives}.

\subsubsection{Structure: Composition of multiples to communicate.}
\label{sec:ds-goals-structure} 

Multiple decision points, alternatives, and outcomes relate to each other with differing structures.

\vspace{0.5em}
\noindent\textbf{Hierarchical decisions.} Multiples can be structured hierarchically, such that early decisions impact the options available for later decisions. For example, tree-of-thought approaches involve sequential steps~\cite{zd} and including a categorical variable in a data analysis pipeline limits the set of usable statistical models~\cite{zp,z1n}. 

\vspace{0.5em}
\noindent\textbf{Independent decisions.} Decisions may also be independent. For example, model features can be independently manipulated for counterfactuals~\cite{zw, zq, zl, z4, zk}, and independent changes to agent personas can generate different outputs~\cite{zj,z2,zx,z1g}. 

\vspace{0.5em}
\noindent\textbf{Outcome set.} Multiples can also be a set of outcomes without decision points or alternatives, such as story ideas~\cite{zh}, analogies or related words~\cite{z1l}, conceptual associations~\cite{zs}, or metaphorical mappings in creative writing~\cite{z1m}.

\vspace{0.5em}
\noindent\textit{Connections.} \textit{Hierarchical} and \textit{independent} decisions can use \textbf{visualizations} with \textit{dependencies} to show their impact on outputs. \textit{Independent decisions} and \textit{outcome sets} may be \textbf{visualized} as \textit{collections}, with decision points as an \textit{elicited details} \textbf{interaction} for \textit{independent decisions}.
%

\subsection{Initiative}
\label{sec:ds-initative} 
In systems that display multiples, different actors (\ref{sec:ds-initative-actor}) take initiative at different times (\ref{sec:ds-initative-timing}). 

\subsubsection{Actor: Who takes initiative.}
\label{sec:ds-initative-actor} 

Systems generate multiples of some granularity at some frequency. They may enable selection of or requesting more multiples.
These steps can be performed by the \textit{user}, the \textit{agent}, as \textit{mixed-initiative} of the user and the agent, or as \textit{inherent} to the system. Initiatives may differ for decision points, alternatives, and outputs (e.g., a user generates the multiverse alternatives and the agent generates the outcome).

\vspace{0.5em}
\noindent\textbf{Generating multiples.} 
The agent typically generates the multiples~\cite{zs,z1c,zv,zr,zp}, but the user may also do so~\cite{zh}. In mixed-initiative settings, the agent generates the decision points or alternatives for the user to then edit~\cite{z11,zq,z1g,za,zv,z1n}. Alternatively, the user may create decision points or alternatives, while the agent generates outcomes~\cite{zp,z1n,z3,z12,z1a,ze}.  %

\vspace{0.5em}
\noindent\textbf{Selecting multiples.} 
Most often, users select from a presented set of decision points or alternatives~\cite{z7,zz,zx,zy,zv,z7}, as well as selecting one from a list of presented outcomes or paths~\cite{z1e,zs,zt,zm,z1j,z15}. Agents may select a preferred option among the multiples to present~\cite{z2,zd}.

\vspace{0.5em}
\noindent\textbf{Requesting more.} 
If the provided multiples are insufficient, agents can call for user input and generation~\cite{z14,zm,z18,z7,zj,zi}, or users can request additional generation from the agent~\cite{z8,zg,z19,z15,z17,z1j}.

\vspace{0.5em}
\noindent\textbf{Determining granularity.} 
The granularity of items to communicate, such as how specific a decision point should be, can be defined by the user~\cite{zd,z3,zh,z1a,zf,z16} or agent~\cite{z5,zu,z1h,z1g}, or inherently as defined by the designer~\cite{z11} or input data~\cite{z1i,z4,z9,zm}. 

\vspace{0.5em}
\noindent\textbf{Determining quantity.} 
The quantity of items to communicate can be defined by the user~\cite{z1a,zh,z1k,zd} or the agent~\cite{z1b,z1h}, or inherently defined by the designer~\cite{zy,z8,z15,z9}.  

\vspace{0.5em}
\noindent\textit{Connections.} 
An \textit{agent's selection} of a preferred option is displayed by \textit{preferred choice} \textbf{explanations}.
The \textit{user} can \textit{generate} or \textit{select} multiples with \textbf{interactions} to \textit{integrate user perspectives} in \textit{iterative}, \textit{linear}, or \textit{on demand} \textbf{initiative timing}. \textit{Requesting more} is associated with \textit{on demand} \textbf{initiative timing}, triggered by \textbf{interactions} to \textit{request more}. 
\textit{Navigation}, \textit{eliciting details}, and \textit{aggregation} \textbf{interactions} help the \textit{user} adjust the \textit{granularity} and \textit{quantity} of information.

\subsubsection{Timing: When initiative takes place.}
\label{sec:ds-initative-timing} 

Generation, selection, and requesting additional multiples can occur at different points of time in a workflow.

\vspace{0.5em}
\noindent\textbf{Once.} 
The system returns a display of multiples all at once~\cite{z12,zp,z1m,z1g,z1d}. The user may optionally interact with the output display, such as through filtering and editing~\cite{z15,ze,z1a,zb,z1e,z1}.

\vspace{0.5em}
\noindent\textbf{Iterative sequence.} 
The agent and user iterate on the same task set, such as by iteratively circling through convergent and divergent stages~\cite{z5,z1k,z17}.

\vspace{0.5em}
\noindent\textbf{Linear sequence.} 
There is a progression through a sequence of steps during which initiative shifts between the user and the agent~\cite{zd,zm,z1m}.

\vspace{0.5em}
\noindent\textbf{On demand.} 
Users can explicitly invoke alternatives on-the-fly as needed~\cite{zq,z19,z17,z16,zk,z10}. Similarly, options can automatically compute and surface as a user adjusts the UI~\cite{zv,z13,zb,z10,z18}. The agent may also monitor the user's task and suggest multiples when relevant~\cite{z18,z14,zm,z18,zi,z10}.

\vspace{0.5em}
\noindent\textit{Connections.}
Including \textbf{content} \textit{up to the current step} best supports an \textit{iterative} or \textit{linear} progression of initiative.
The \textbf{interactions} of \textit{requesting more} and \textit{integrating perspectives} support \textit{iterative}, \textit{linear}, and \textit{on demand} progressions of initiative, while \textit{navigation}, \textit{eliciting details}, and \textit{aggregation} are more often used to explore a display generated \textit{once}.

\subsection{Content}
\label{sec:ds-content} 
Systems can display varied quantities of the multiples' content to the user (\ref{sec:ds-content-quantity}),  accompanied by additional explanatory content (\ref{sec:ds-content-expl}).

\subsubsection{Quantity: What multiples to initially display.}
\label{sec:ds-content-quantity} 

Systems display various subsets of the multiples available.

\vspace{0.5em}
\noindent\textbf{All.} 
The system displays all options~\cite{zr,zl,zj}. 

\vspace{0.5em}
\noindent\textbf{Samples.} The system displays a subset of one or more multiples~\cite{zb,zu,zc}, sometimes continuously adding more as they are computed~\cite{z3}.

\vspace{0.5em}
\noindent\textbf{Current step.} 
The system progressively reveals alternatives for the current step as the user advances through decision points ~\cite{zd,z14,zj,z1e}.

\vspace{0.5em}
\noindent\textbf{No outcomes.} 
The system displays only the decision points and alternatives, such that user interaction is required to then show outcomes~\cite{z1k, zv}. 

\vspace{0.5em}
\noindent\textit{Connections.}
A continuously updating \textit{sample} of options while computing can be paired with \textbf{explanatory} \textit{performance metrics} to reveal bugs or converging outcomes. Displaying \textit{all} multiples can be paired with \textbf{explanatory} \textit{aggregations} to improve interpretation despite a large scale. 
If only \textit{decision points} are displayed initially, users may use \textit{navigation} \textbf{interactions} to filter by decision points.

\subsubsection{Explanations: What supplemental content to show.}
\label{sec:ds-content-expl} 
Systems may provide additional explanatory content alongside the main multiples information. 

\vspace{0.5em}
\noindent\textbf{Preferred choice.} 
The system can indicate an option deemed better than the alternatives using natural language~\cite{z1b,z2,z14,z1e}, color highlighting~\cite{zd,zn,z1h,zr,zy}, or path ranking~\cite{zd,zr,z13,z10,z12,z4}. 

\vspace{0.5em}
\noindent\textbf{Path details.}
Systems can provide additional descriptions of and rationales for path choices~\cite{zv,z1m,zj}.

\vspace{0.5em}
\noindent\textbf{Difference highlighting.} 
Displays can emphasize differences through color~\cite{zi,za,zn,z1c,z1l,z4}, text~\cite{z1d}, clustering~\cite{z1l,z3,z4,zd}, visual indicators like arrows~\cite{z1i,zw}, and size and position~\cite{zc,z1l,z4,zv}.
Color~\cite{z1m,z1g,zo} and sorting~\cite{z1i} may also be used to support navigating large amounts of content.

\vspace{0.5em}
\noindent\textbf{Decision sensitivity.} 
The sensitivity of different decision points on the outcomes can be shown with quantitative values~\cite{zn,z1h} or color~\cite{z4,zk,zp,z3,z12}.

\vspace{0.5em}
\noindent\textbf{Performance metrics.} 
Systems can communicate errors or path quality with color~\cite{za,zb,z3,zo} or physical separation~\cite{z3}. Designers may provide additional performance metrics, such as runtime warnings and summaries of model errors~\cite{z3,zo}.

\vspace{0.5em}
\noindent\textbf{Trend aggregations.} 
Aggregated data can communicate trends and overviews~\cite{z13,zq,z1a,ze,z1,zo}. For instance, multiverse plots bin the frequency of decision points for a range of outcomes~\cite{zn}. However, aggregation can be misleading, such as by conflating the probabilistic and possibilistic uncertainty of running many analyses~\cite{ze}.

\vspace{0.5em}
\noindent\textit{Connections.}
\textit{Performance metrics}, \textit{decision sensitivity}, and \textit{highlighting differences} are generally used by \textbf{visualizations} \textit{with dependencies}. 
If displays begin with \textit{aggregation}, users can request \textit{details on demand}.

\subsection{Display}
\label{sec:ds-display} 

The final display shows the visualization (\ref{sec:ds-display-vis}) and enables user interactions (\ref{sec:ds-display-interactions}).

\subsubsection{Visualization: How to display multiples.}
\label{sec:ds-display-vis} 
Designers can visualize multiples as collections or with dependencies.

\vspace{0.5em}
\noindent\textbf{As collections.} 
Unstructured sets of multiples without decision points visible can be displayed as a moodboard-like canvas~\cite{zh,z1m,z19,z13,z5} or a text list~\cite{z8,z1j,z2,z1e,z10}. Other less common displays include a radial visualization~\cite{zc}, word cloud~\cite{z1l}, or 3D environments~\cite{zx,zy}.

\vspace{0.5em}
\noindent\textbf{Dependency displays.} 
Tree and graph visualizations can display decision points, alternative paths, and their outcomes~\cite{z14,za,z16,zs,z1n,zg}. 
Node-link diagrams, pipe-diagrams, and icicle plots also display dependent paths and enable comparison of quality among them~\cite{zf,zd,zo}.
Other displays include paths as vertical straight sequences listed side by side~\cite{z15,zb}, a decision matrix in a specification curve design~\cite{ze,zn,z1a}, superimposed entries to a radar chart visualization~\cite{zl}, a Sankey diagram~\cite{zq}, a table~\cite{z1d}, or plain text~\cite{z11,zo,z1b}. 

Outputs for these displays can be raw values~\cite{zo,z1c}, natural language text~\cite{zl,z1c}, a value on a slider~\cite{zo,zw,z1c}, or displayed graphically as a density plot~\cite{zn,zp}, dot plot~\cite{z3}, or line plot~\cite{z1}. With just two selected decision points of interest, scatter plots~\cite{z9,z1k} or partial dependence plots~\cite{z1h} put plot outcomes along decision points as axes. 

\vspace{0.5em}
\noindent\textit{Connections.}
\textit{Collection} displays may enable \textit{detail elicitation} \textbf{interactions} to see input decisions. \textbf{Interactions} to \textit{integrate perspectives} by comparing among decision trees can be supported by node link, pipe diagram, and icicle plot variations on \textit{tree} displays.

\subsubsection{Interactions: How users can interface with the display.}
\label{sec:ds-display-interactions} 
Interactions allow users to engage with the display.

\vspace{0.5em}
\noindent\textbf{Integrate perspectives.} Users can choose preferred~\cite{z18,z1n,zz,zx,z8}, rank~\cite{zh,zc,ze}, delete or deprioritize unwanted multiple(s)~\cite{ze,zv,z18,z1n,zh}.
Users can also edit~\cite{zh,zg,z1g,z8}, add~\cite{zd}, add explanations to~\cite{z1n}, split~\cite{zo}, or combine~\cite{zh,z1n,z19} multiples. Changed mid-level decisions may then re-execute~\cite{zg,z1g}.

\vspace{0.5em}
\noindent\textbf{Navigate.} 
Users may navigate large swaths of information by filtering~\cite{zr,z1a,z1}, selecting multiples of interest from given options~\cite{z5,z1k,z1i,zv,z9,zp}, brushing and linking~\cite{z3,ze,z4,zf}, and filtering subsets via sliders~\cite{z1i}.
Users can also navigate by sorting~\cite{ze}, searching via a text box~\cite{z1k}, or switching between multiple views~\cite{zp,zr,z3}.

\vspace{0.5em}
\noindent\textbf{Aggregate.} 
To reduce detail, users may create their own aggregations by zooming out~\cite{ze,zd,z1m}, manipulating the granularity of multiples~\cite{z1k,za}, or grouping multiples into higher-level conceptual categories~\cite{zq,zv}. 

\vspace{0.5em}
\noindent\textbf{Elicit details.} 
Users may zoom into the canvas~\cite{zd,zc,z1k} or interact with multiples~\cite{z17,zc,z5,zj,z1i,zb} to elicit additional detail. Users may also elicit details or explanations from a model~\cite{zi,zd,zv,z15,z1g}. 

\vspace{0.5em}
\noindent\textbf{Request more.} 
Users can explicitly call for agent generation through the interface~\cite{zv,z1k,zi,z1j,z8}. Alternatively, edits and interactions may automatically call the agent~\cite{zg,z1j,z19,zh,z15,zb}. Users can provide additional information to steer the generation~\cite{z1m,z10,z17,z1j,zz,zk}. Displays enabling these calls may also enable return to previous states with an undo button~\cite{zh} or displays of the history of edits~\cite{z15,z19}.

\subsection{Utilizing the design space}
\label{sec:ds-utilizing} 
We exemplify how the design space can aid the generation of Agentic Path Maps and empirical study with our user study design and stimulus.
Empirical studies can refine our understanding of how decisions along the design space dimensions impact different \textbf{objectives}. 
We orient our study and stimulus design to support the \textit{improving understanding} \textbf{objective} of a diverging decision space. We manipulate the stimulus along the \textbf{quantity} of content shown dimension since we predict Agentic Path Maps will increase computational and cognitive costs. We aim for simplicity to focus on this dimension of interest.

With our objective defined, we progress through each design space dimension to design the stimulus. Our user study uses a data analysis agent. Analysis decisions impact one another, so we must communicate information with a \textit{hierarchical} \textbf{structure}. 
We next consider \textbf{initiatives} in the system. We will tell users that the content is \textit{generated} by an \textit{agent}, although for the purposes of controlling the study, the \textit{generation}, \textit{granularity}, and \textit{quantity} will instead be determined by the \textit{designer}. For simplicity, we will not have initiative for \textit{selection} or \textit{calling for more generation}. Improving user understanding most often aligns with \textbf{timing initiative} to generate \textit{once} and interact after. 

We manipulate the \textbf{quantity} dimension to show \textbf{all} analyses or one or three \textbf{samples}. Additional \textbf{explanatory} content aids understanding, so we include \textit{path details} and \textit{decision point sensitivity}. 
As we are displaying \textbf{hierarchical} content, we refer to \textbf{visualizations} \textbf{with dependencies}, selecting the specification curve as it has prior grounding in the data analysis domain and can easily be manipulated to show samples. Finally, we enable a \textbf{details on demand} interaction to show \textit{path details} and support \textit{improved understanding}. We describe the final interface in Section~\ref{sec:user-interfaces-treatment}.

\begin{figure*}
  \centering
  \includegraphics[width=\linewidth]{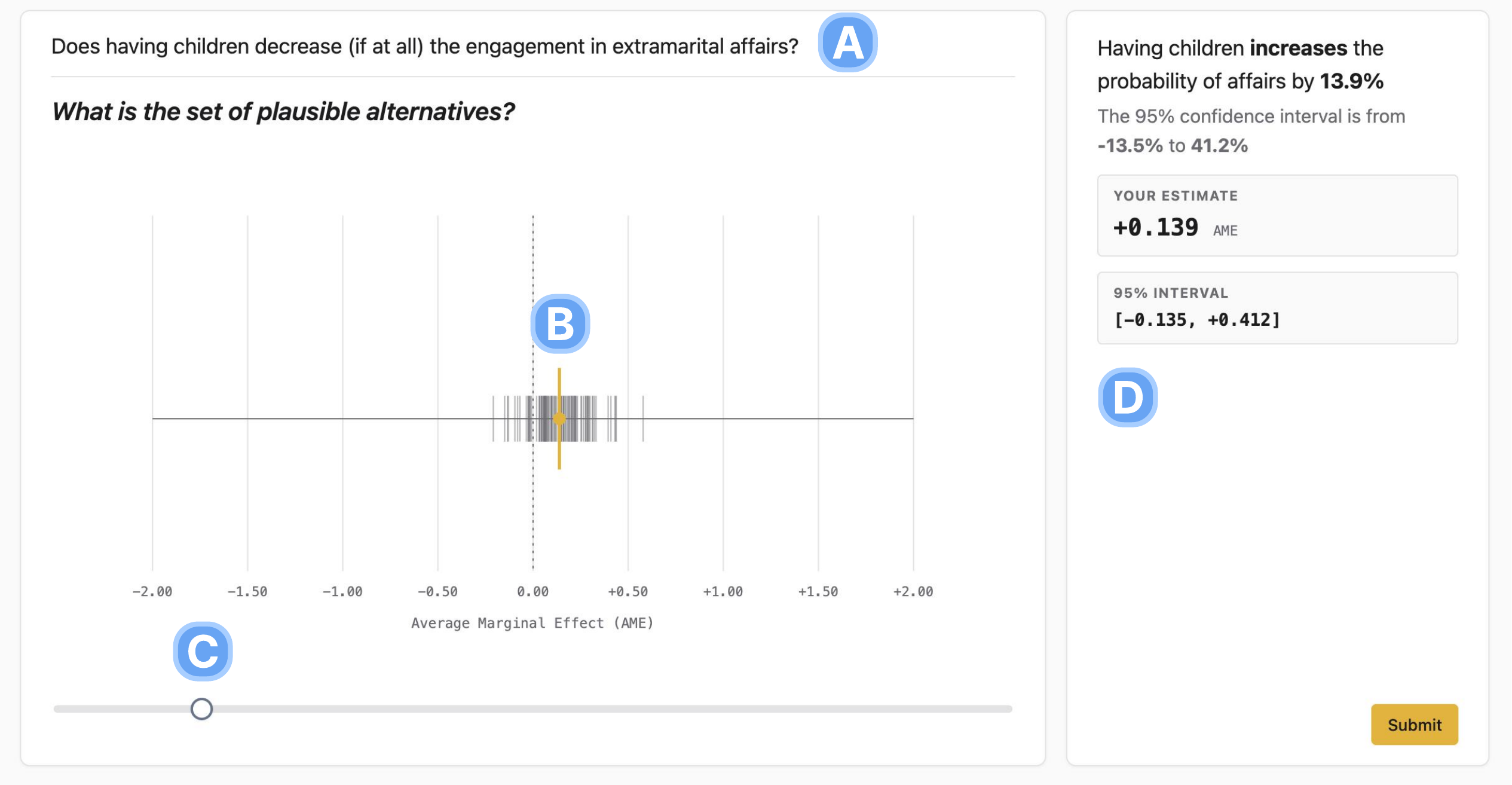}
  \caption{Belief elicitation figure. Participants reviewed the analysis question [A] and input what they thought was the ``most likely relationship'' by moving the yellow bar [B]. They then used a slider [C] to adjust a distribution of grey bars indicating the ``set of plausible alternatives.'' The right panel contains the exact values and an interpretive sentence [D].}
  \Description{A belief-elicitation screen with a wide plot panel on the left and a narrower summary panel on the right. The horizontal axis shows average marginal effect ranging from negative to positive two with a dotted reference line at 0. The selected estimate is 0.139, emphasized with yellow, with possible values distributed on both sides of the origin. The summary reports a 95 percent confidence interval ranging from negative 0.135 to (positive) 0.412.}
 \label{fig:elicitation}
\end{figure*}

\section{User study}
\label{sec:user}

We study Agentic Path Maps applied to the data analysis domain (\ref{sec:user-background}). We generate data with an AI agent (\ref{sec:user-data}) to display on interfaces showing one or multiple possibilities (\ref{sec:user-interfaces}). After recruiting participants (\ref{sec:user-participants}) we elicit their beliefs before and after viewing the interfaces (\ref{sec:user-procedure}). We measure variables related to their beliefs, duration of interaction, and cognitive load (\ref{sec:user-variables}) and test them with preregistered hypotheses (\ref{sec:user-hypotheses}) and models (\ref{sec:user-methods}). We provide details on data generation, study materials, and modeling code in Supplementary Materials.

\subsection{Data analysis application}
\label{sec:user-background} 

Data analysis agents, such as DataVoyager~\cite{ai22025asta}, allow users to input queries with data and return analysis code and a result. 
The data analysis domain provides useful grounding due to its quantitative outcomes and prior work on multiverse analyses. 
Multiverse analyses consider multiple traversals through the ``garden of forking paths'' in tandem~\cite{gelman2013garden,simonsohn2015specification}.
Rather than replacing bespoke interfaces~\cite{sarma2024milliways,liu2021boba,z3}, we use multiverses to test Agentic Path Maps.

\subsection{Data generation}
\label{sec:user-data} 
To ensure the displayed data is task-valid, we use multiverse specifications created by expert analysts in the BLADE benchmark~\cite{gu2024blade}. We filter to datasets with results spanning from positive to negative: ``affairs,'' which measures if having children affects the rate of extramarital affairs, and ``crofoot,'' which measures if the size and location of capuchin monkey fights affect the likelihood of winning, for which we use the size coefficient. We use all nine remaining analyses from affairs and randomly select from crofoot to match, while ensuring the median, minimum, and maximum were included. To present results intuitively, we use average marginal effects for affairs and odds ratios with z-scores for crofoot. 
We prompt DataVoyager to complete each analysis's specification and extract and verify DataVoyager's output code by hand.
We generate specifications and rationales using Claude Sonnet 5~\cite{claude-sonnet-5} with default reasoning settings.

\begin{figure*}
  \centering
  \includegraphics[width=\linewidth]{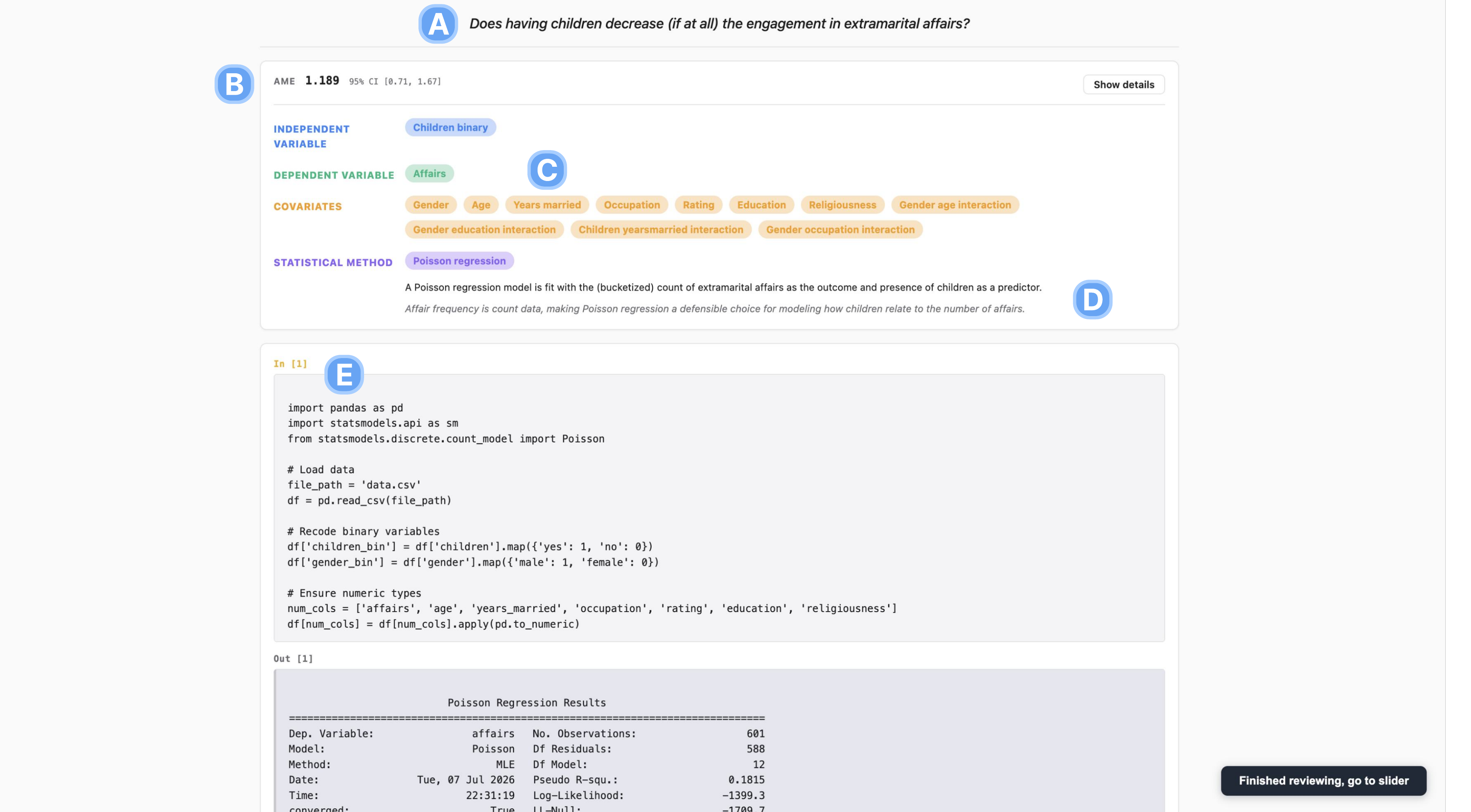}
  \caption{Baseline interface displaying one analysis. The interface shows the [A] analysis question. [B] found output, [C] what analysis decisions were made, [D] specifications and rationales, as well as [E] code.}
  
  \Description{
  A single-analysis interface with a result card above Jupyter notebook-style code cells. The question at the top, marked A, asks whether having children decreases engagement in extramarital affairs. The result, marked B, is an average marginal effect of 1.189 with a 95 percent confidence interval from 0.71 to 1.67. Below it, chips marked with the label C identify Children binary as the independent variable, Affairs as the dependent variable, covariates, and a Poisson regression as the method. Beneath the method chip a label marked D describes the model and gives an italicized rationale. The notebook below, marked with the label E, shows Python code and the resulting Poisson regression summary.
  }
 \label{fig:baseline}
\end{figure*}

\subsection{Interfaces}
\label{sec:user-interfaces}

We design a baseline and treatment interface to display one or multiple analysis path(s), and an interface to elicit beliefs.
We used Claude Code~\cite{claude-code} while developing. 

\subsubsection{Baseline interface: displaying one path}
The baseline interface (Figure~\ref{fig:baseline}) displays the analysis question [A] and DataVoyager's result [B] near the top. It summarizes the analysis decision categories (e.g., statistical method) and displays buttons for choices made [C]. Clicking buttons, or the ``Show details'' button, reveals a short specification and rationale [D]. DataVoyager's code displays at the bottom in notebook format [E].

\subsubsection{Treatment interface: displaying multiple paths}
\label{sec:user-interfaces-treatment}

\begin{figure*}
  \centering
  \includegraphics[width=\linewidth]{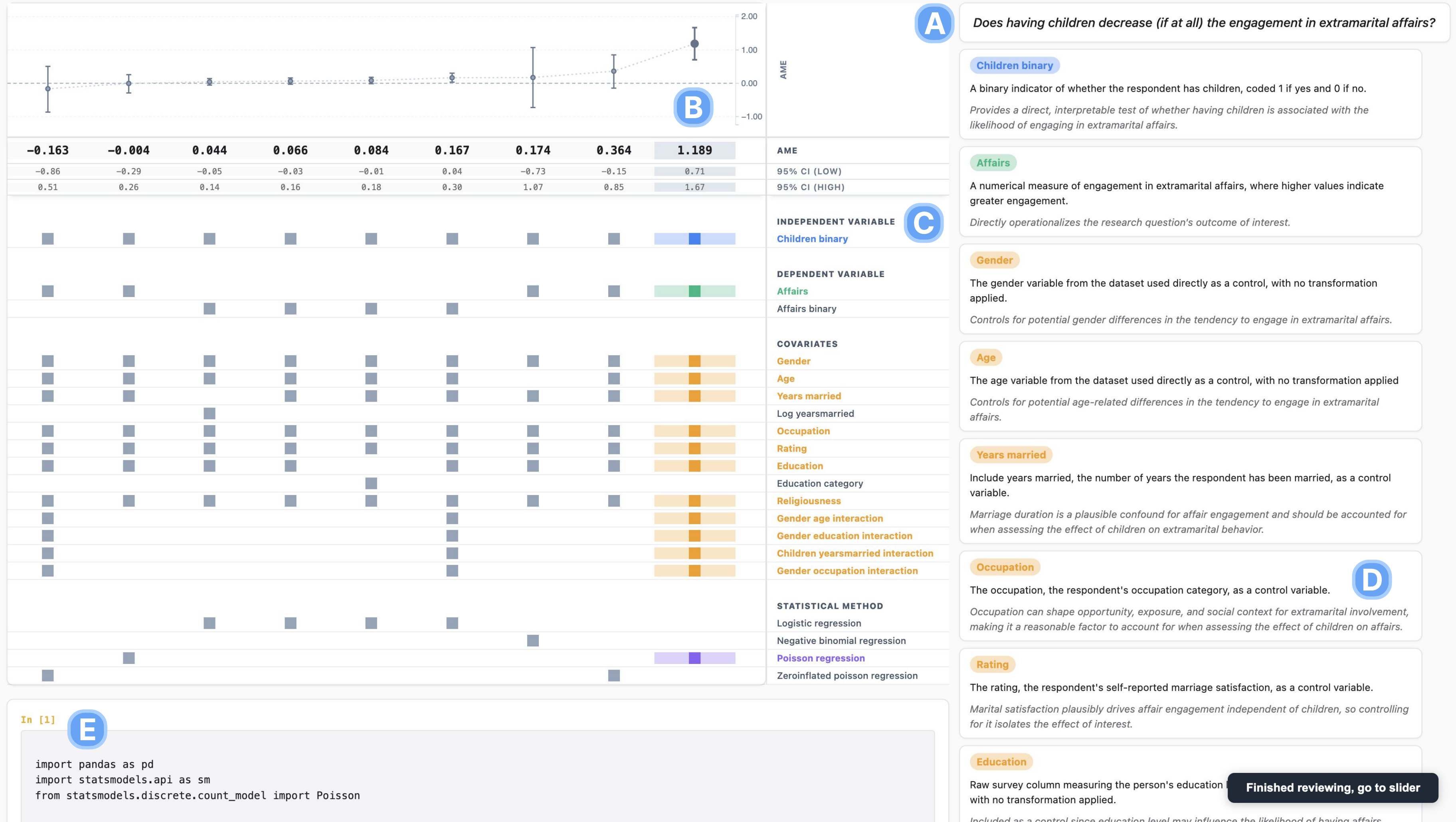}
  \caption{Treatment interface displaying multiple analyses. This Agentic Path Maps display of multiple analysis alternatives uses a specification curve \textit{dependency} \textbf{visualization}. Every row indicates a defensible analysis choice, and every column an individual analysis with associated code and an output. The boxes indicate which combination of alternative analysis choices was used for that column's analysis. The display shows the [A] analysis question, [B] found outputs, [C] what analysis decisions were made, [D] specifications and rationales, and [E] code. }
  
  \Description{A multiple-analysis interface with a plot and choice matrix on the left and explanation cards on the right. The question at the top right is marked A. A specification curve marked B orders nine average marginal effects from negative 0.163 to 1.189, with vertical confidence intervals. Three rows below give estimates and interval bounds. In the matrix marked C, columns align with the plotted analyses and rows list choices for variables, covariates, and statistical methods. Squares indicate the choices used in each analysis. The rightmost column is highlighted, linking its choices to cards marked D that provide specifications and italicized rationales. A notebook cell marked E sits below and displays the selected analysis's Python code.}
 \label{fig:treatment}
\end{figure*}

The treatment interface (Figure~\ref{fig:treatment}) adopts a specification curve \textit{dependency} \textbf{visualization} from the design space~\cite{simonsohn2015specification}. It displays the analysis question [A] and multiple outputs with raw and plotted numbers sorted from smallest to largest [B]. The rows of a decision matrix communicate decision points like ``covariates'' (sets of rows) and alternatives like ``gender'' (individual rows) [C]. Each column represents a different analysis path, with boxes indicating if an analysis used the alternative in that row. Clicking on a column, individual row header, or box reveals cards with specifications and rationales [D]. Clicking on a decision point header reveals a rationale describing the tradeoffs among that decision point's alternatives. The code for a selected column appears at the bottom [E].

\subsubsection{Belief elicitation.}
We elicit participants' beliefs about an analysis question [A] before and after viewing the output (Figure~\ref{fig:elicitation}).
We use a variant of the Line + cone approach by Karduni et al.~\cite{karduni2021bayesian}, after consulting a variety of approaches~\cite{goldstein2014lay,xiao2026priorweaver,zondervan2017application,johnson2010methods,xiong2022seeing}. Participants place a yellow line [B] on an axis representing the ``most likely relationship,'' then use a slider [C] to adjust the width of 100 gray lines drawn from a normal distribution to indicate the ``set of plausible alternatives.'' They view exact numbers and an interpretive sentence on the right panel [D].

\subsection{Participant recruitment and assignment}
\label{sec:user-participants}

Participant recruitment and the procedure were approved by our university's Institutional Review Board. 
We recruited novice analysts who would be familiar with statistical methods but lack the expertise to create a multiverse analysis themselves. We recruited participants within our university's message boards and participant pool, as well as through our networks. Our interest form requested self-reported familiarity with statistical methods and Python on a 1-5 scale. We moved forward with those who self-reported as a 3-4 on statistical methods and a 2-5 on Python.  
Participant assignments (16 cells, Table~\ref{tab:participants}) followed a mixed design: the tasks and main conditions were within-subjects, and sub-conditions were between-subjects~\cite{bielicke2025planet}. We recruited 48 participants, the most feasible within our resource constraints.

\subsection{Procedure}
\label{sec:user-procedure}

Sessions were conducted for one hour over Zoom and compensated with a \$50 gift card. Participants signed a consent form, confirmed they were unfamiliar with the datasets, read a text description of the study, and reviewed a practice elicitation example.

For each task, participants reviewed a summary of the dataset and relevant outcome measure. We then elicited participants' prior beliefs. Next, participants viewed a practice example and answered a series of comprehension questions. They then viewed the results, which we referred to as AI-generated. We instructed participants to proceed when they felt ready, encouraging them to do so after ten minutes. We then elicited their posterior beliefs. They could not see the outcomes while reporting their posterior. 
Participants next completed a Cognitive Load questionnaire~\cite{krieglstein2023development} and reflected on what they would do next.

After completing both tasks, participants completed an outtake questionnaire indicating their preference between interfaces, AI application use, and trust in AI~\cite{mcgrath2025measuring}. We asked participants for open-ended reflection on the interfaces and elements they would change. 
Finally, we disclosed that the multiverse exists for both tasks and that analyses came from a human-made benchmark.

\subsection{Variables}
\label{sec:user-variables}

We manipulate the number of analyses shown and measure outcomes related to elicited beliefs, duration of interaction, and cognitive load.

\subsubsection{Manipulated}
\label{sec:user-variables-manipulated}

For our preregistered hypotheses, we manipulate whether participants see one or multiple analyses. For the baseline interface, we show the analysis with the minimum or maximum output. Using the outer results enables us to isolate the behavior of anchoring to one task-valid result to establish if an effect exists. 
Our non-preregistered analyses address sub-conditions. Baseline sub-conditions display values that agree or disagree with the polarity of participants' initial beliefs. Treatment sub-conditions display a sample of three (the minimum, median, and maximum outputs), or all nine analyses. 

\subsubsection{Non-manipulated}
\label{sec:user-variables-nonmanipulated}

Our quantitative metrics refer to the task-valid set as a whole. Although there is no universally ``correct'' takeaway from a multiverse~\cite{sarma2024milliways,del2021traveler}, we consider heavily updating away from the multiverse median towards an extreme value to be a misleading takeaway.
Weight of Advice (WoA) represents how heavily humans weigh displayed information against their prior. We use Weight of Advice, calculated as 
((final\_value - initial\_value) / (displayed\_value - initial\_value))~\cite{bailey2023meta}, as a measurement of overreliance, using z-scores for consistency across tasks~\cite{ibrahim2025measuring,passi2025addressing}. In line with Bayesian reasoning, we use the value towards which the participant updates as the displayed\_value for the treatment condition. WoA outputs should be from 0 (ignore likelihood) to 1 (ignore prior) with rational behavior~\cite{bailey2023meta}. Our pilot data reflected that range, but the user study had large outliers, so we deviated from the preregistration by removing values lower than -0.5 or higher than 1.5.

We measure the absolute distance of the user's posterior beliefs to the multiverse median using z-scores, the width of the posterior's confidence interval normalized by the slider's range, and the number of seconds users spent reviewing the outputs.  
We measure cognitive load using a validated questionnaire from Krieglstein et al.~\cite{krieglstein2023development}. They distinguish among three types of Cognitive Load: Intrinsic (caused by task complexity), Extraneous (caused by presentation complexity), and Germane (productive cognitive effort involved in learning).\footnote{Germane Cognitive Load is a newer concept than Intrinsic and Extraneous Cognitive Loads, and the theory is still being refined by cognitive scientists. Krieglstein et al. include it in the finalized validated questionnaire.} Each score averages 5 items with a 1-9 scale.

\begin{figure*}
  \centering
  \includegraphics[width=\linewidth]{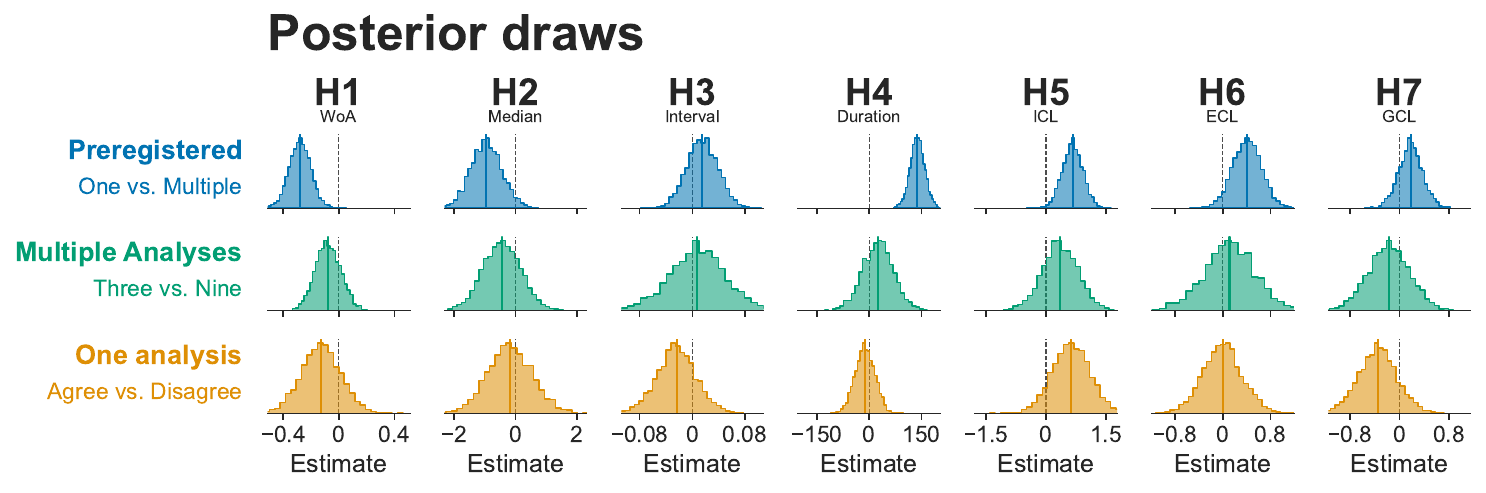}
  \caption{Posterior draws. We report the posterior draws from our Bayesian linear mixed models for each of the preregistered hypotheses and sub-condition analyses. Showing nine analyses compared to three resulted in changes in the same direction as showing multiple analyses compared to one, with the exception of Germane Cognitive Load (H7).}
  \Description{21 posterior density plots are arranged in three rows and seven columns. The first row is labeled Preregistered with the subheader One vs. Multiple, the second with Multiple analyses with a Three vs. Nine subheader, and the third with One analysis with an Agree vs Disagree subheader, from top to bottom respectively. Columns correspond to hypotheses H1 through H7. Rows show the main posterior distributions, treatment estimates, and baseline estimates. Each plot shows the distribution of posterior draws along an estimate axis and includes a dashed vertical marker at zero. Axis ranges differ across hypotheses. In the Preregistered row, H1 and H2 are centered below zero, H3 is centered near zero, and H4 through H7 are centered above zero; H4 has a much larger numerical scale than the other hypotheses.}
 \label{fig:posterior_draws}
\end{figure*}

\subsection{Hypotheses}
\label{sec:user-hypotheses}

We preregistered\footnote{\url{https://osf.io/wvyef/overview?view_only=a6981669ee7a4b0197edd3a381af2249}} seven hypotheses that are motivated by Bayesian models of cognition~\cite{griffiths2008bayesian}, in which human prior beliefs update towards viewed data, anchoring bias~\cite{tversky1974judgment}, in which humans anchor on initial seen outputs, and Cognitive Load theory, in which showing more complex learning materials increases burden on the user~\cite{sweller2011cognitive}.

\vspace{0.5em}
\noindent\textbf{[H1]} Participants' Weight of Advice on an extreme output is lower when shown multiple paths than when shown one extreme path.

\vspace{0.5em}
\noindent\textbf{[H2]} Participants' posterior beliefs are closer to the multiverse median when shown multiple paths than when shown one extreme path.

\vspace{0.5em}
\noindent\textbf{[H3]} Participants' posterior confidence interval is wider when shown multiple paths than when shown one extreme path. 

\vspace{0.5em}
\noindent\textbf{[H4]} Participants spend more time reviewing when shown multiple paths than when shown one extreme path.

\vspace{0.5em}
\noindent\textbf{[H5]} Participants' Intrinsic Cognitive Load is higher when shown multiple paths than when shown one extreme path. 

\vspace{0.5em}
\noindent\textbf{[H6]} Participants' Extraneous Cognitive Load is higher when shown multiple paths than when shown one extreme path. 

\vspace{0.5em}
\noindent\textbf{[H7]} Participants' Germane Cognitive Load is higher when shown multiple paths than when shown one extreme path.

\begin{table*}[t]
\centering
\caption{Results for preregistered hypotheses. We find a large or very large effect with a high posterior probability (Post. Prob) that participants place a lower Weight of Advice on extreme values [H1] and end with beliefs closer to the multiverse median [H2] when shown alternative analysis pathways, but report higher Intrinsic [H5] and Extraneous Cognitive Load [H6]. Participants interacted for longer [H4] and reported higher beneficial Germane Cognitive Load with a small effect [H7].}
\Description{Seven hypothesis rows with seven columns: hypothesis (abbreviated as "hypo."), Description, standardized effect (abbreviated as Effect), the Effect Confidence interval, estimate in the outcome's original units, its confidence interval, and posterior probability of the hypothesized direction. Standardized effect intervals exclude 0 for H1, H4, and H5. H2's upper bound rounds to zero. Estimate intervals exclude zero for H1, H2, H4, H5, and H6. H3 and H7 both have intervals spanning zero.}
\label{tab:main_results}
\begin{tabular}{llrrrrr}
Hypo. & Description & Effect  & Effect CI            & Estimate   & Estimate CI              & Post. Prob \\ \hline
H1 & Lower WoA   & -0.63 & {[}-0.99, -0.24{]} & -0.28  & {[}-0.41, -0.15{]}   & 0.9992     \\
H2 & Closer to median  & -0.38 & {[}-0.76, 0.00{]}   & -0.95  & {[}-1.78, -0.10{]}   & 0.9645     \\
H3 & Larger interval & 0.09  & {[}-0.23, 0.39{]}  & 0.02   & {[}-0.02, 0.05{]}    & 0.7385     \\
H4 & Longer duration  & 0.96  & {[}0.68, 1.25{]}   & 137.94 & {[}103.30, 172.65{]} & 1.0000     \\
H5 & Higher ICL  & 0.40  & {[}0.10, 0.70{]}   & 0.67   & {[}0.24, 1.09{]}     & 0.9935     \\
H6 & Higher ECL  & 0.31  & {[}-0.06, 0.67{]}  & 0.41   & {[}0.01, 0.80{]}     & 0.9522     \\
H7 & Higher GCL  & 0.17  & {[}-0.17, 0.50{]}  & 0.18   & {[}-0.15, 0.51{]}    & 0.8238    
\end{tabular}
\end{table*}

\subsection{Statistical methods}
\label{sec:user-methods}

\begin{sloppypar}
For all hypotheses, we use a Bayesian linear mixed model with task as a fixed effect and participant as a random effect: $dependent\_variable \sim condition + task + (1 \mid pid)$. We remove participant in sub-condition analyses and use the brms package (2.23.0) in R (4.6.1). We create weakly informed priors using pilot data, as there is no previous research on this topic~\cite{phelan2019some}. We used Claude~\cite{claude-sonnet-5} for assistance with interpreting jargon, data wrangling, and checking model code.
\end{sloppypar}

Bayesian statistics align with HCI methods and reduce the likelihood of inflating effect sizes with small $n$ studies~\cite{kay2016researcher}. Rather than dichotomous inference~\cite{besancon2019continued}, we report results from the brms `hypothesis()` and `summary()` functions as well as with default settings from the effectsize package (1.0.3) with default parameters. We do not adjust for multiple hypotheses as we do not use dichotomous inference. We use Funder et al.'s effect size thresholds for psychological research of 0.05 for very small, 0.1 for small, 0.2 for medium, 0.3 for large, and 0.4 for very large~\cite{funder2019evaluating}.

\begin{table}[t]
\centering
\caption{Participant assignments. We recruit 48 participants such that each of the rows is assigned to three participants. Baseline (Agree) and Baseline (Disagree) are sub-conditions displaying an analysis agreeing or disagreeing with the user's prior. Treatment (Three) and Treatment (Nine) are sub-conditions displaying three or nine analyses.}
\label{tab:participants}
\Description{Sixteen assignment rows with four columns, grouped into two pairs. Each pair gives the interface and task for one trial. Every row pairs one baseline interface with one treatment interface and assigns Affairs in one trial and Crofoot in the other. The rows cover all combinations of baseline agreement or disagreement, three or nine treatment analyses, interface order, and task order. The first eight rows begin with baseline, and the remaining eight begin with treatment. }
\begin{tabular}{ll|ll}
\hline
Trial 1 Interface & Trial 1 Task & Trial 2 Interface & Trial 2 Task \\ \hline
Baseline (Agree)                & Affairs      & Treatment (Three)                & Crofoot      \\
Baseline (Agree)                & Affairs      & Treatment (Nine)                & Crofoot      \\
Baseline (Agree)                & Crofoot      & Treatment (Three)                & Affairs      \\
Baseline (Agree)               & Crofoot      & Treatment (Nine)                & Affairs      \\
Baseline (Disgree)                & Affairs      & Tretment (Three)                & Crofoot      \\
Baseline (Disgree)                & Affairs      & Treatment (Nine)                & Crofoot      \\
Baseline (Disgree)                & Crofoot      & Treatment (Three)                & Affairs      \\
Baseline (Disgree)                & Crofoot      & Treatment (Nine)                & Affairs      \\
Treatment (Three)                & Affairs      & Baseline (Agree)                & Crofoot      \\
Treatment (Three)                & Affairs      & Baseline (Disgree)                & Crofoot      \\
Treatment (Three)                & Crofoot      & Baseline (Agree)                & Affairs      \\
Treatment (Three)                & Crofoot      & Baseline (Disgree)                & Affairs      \\
Treatment (Nine)                & Affairs      & Baseline (Agree)                & Crofoot      \\
Treatment (Nine)                & Affairs      & Baseline (Disgree)                & Crofoot      \\
Treatment (Nine)                & Crofoot      & Baseline (Agree)                & Affairs      \\
Treatment (Nine)                & Crofoot      & Baseline (Disgree)                & Affairs     
\end{tabular}
\end{table}

\begin{table}[t]
\centering
\caption{Sub-condition results. We test the differences between showing three and nine analyses (T-) and showing one outcome that agrees or disagrees with their prior (B-). For the treatment sub-conditions, most effects are in the same direction as the main hypotheses but of a smaller magnitude, except for Germane Cognitive Load (T-H7).}
\label{tab:expl_results}
\Description{Five columns show the hypothesis identifier, standardized effect, its interval, the estimate in the outcome's original units, and its interval. Fourteen rows form two blocks: seven treatment comparisons labeled T-H1 through T-H7, followed by seven baseline comparisons labeled B-H1 through B-H7. Every reported effect interval and estimate interval spans zero. Treatment point estimates are negative for H1, H2, and H7 and positive for H3 through H6. Baseline point estimates are negative except for H5, which is positive, and H6, whose estimate rounds to zero.}
\begin{tabular}{lrrrr}
Expl.        & Effect     & Effect CI              & Estimate     & Estimate CI                \\ \hline
\multicolumn{5}{l}{Treatment: Displaying 3 vs 9 samples}            \\ \hline
T-H1         & -0.20    & {[}-0.74, 0.34{]}    & -0.08    & {[}-0.25, 0.10{]}      \\
T-H2         & -0.24    & {[}-0.80, 0.34{]}    & -0.45    & {[}-1.66, 0.77{]}      \\
T-H3         & 0.03     & {[}-0.40, 0.44{]}    & 0.01     & {[}-0.08, 0.09{]}      \\
T-H4         & 0.18     & {[}-0.42, 0.74{]}    & 25.65    & {[}-64.49, 115.45{]}   \\
T-H5         & 0.22     & {[}-0.36, 0.78{]}    & 0.35     & {[}-0.56, 1.24{]}      \\
T-H6         & 0.08     & {[}-0.47, 0.68{]}    & 0.11     & {[}-0.77, 0.97{]}      \\
T-H7         & -0.14    & {[}-0.75, 0.43{]}    & -0.17    & {[}-0.84, 0.49{]}      \\ \hline
\multicolumn{5}{l}{Baseline: Output agrees vs disagrees with prior} \\ \hline
B-H1         & -0.22    & {[}-0.75, 0.31{]}    & -0.13    & {[}-0.40, 0.15{]}      \\
B-H2         & -0.09    & {[}-0.68, 0.47{]}    & -0.19    & {[}-1.60, 1.24{]}      \\
B-H3         & -0.13    & {[}-0.54, 0.29{]}    & -0.02    & {[}-0.08, 0.04{]}      \\
B-H4         & -0.12    & {[}-0.71, 0.48{]}    & -11.64   & {[}-67.42, 42.60{]}    \\
B-H5         & 0.38     & {[}-0.20, 0.92{]}    & 0.63     & {[}-0.31, 1.58{]}      \\
B-H6         & 0.01     & {[}-0.59, 0.60{]}    & 0.00     & {[}-0.65, 0.65{]}      \\
B-H7         & -0.33    & {[}-0.89, 0.26{]}    & -0.35    & {[}-0.98, 0.28{]}      
\end{tabular}
\end{table}

\section{Results}
\label{sec:results}

\begin{sloppypar}
We report results for our preregistered hypotheses (\ref{sec:results-prereg}), non-preregistered analyses (\ref{sec:results-expl}), and qualitative feedback (\ref{sec:results-qual}). We provide model results and outtake questionnaire responses in Supplementary Materials.
\end{sloppypar}

\subsection{Preregistered hypotheses}
\label{sec:results-prereg} 
\begin{figure*}
  \centering
  \includegraphics[width=\linewidth]{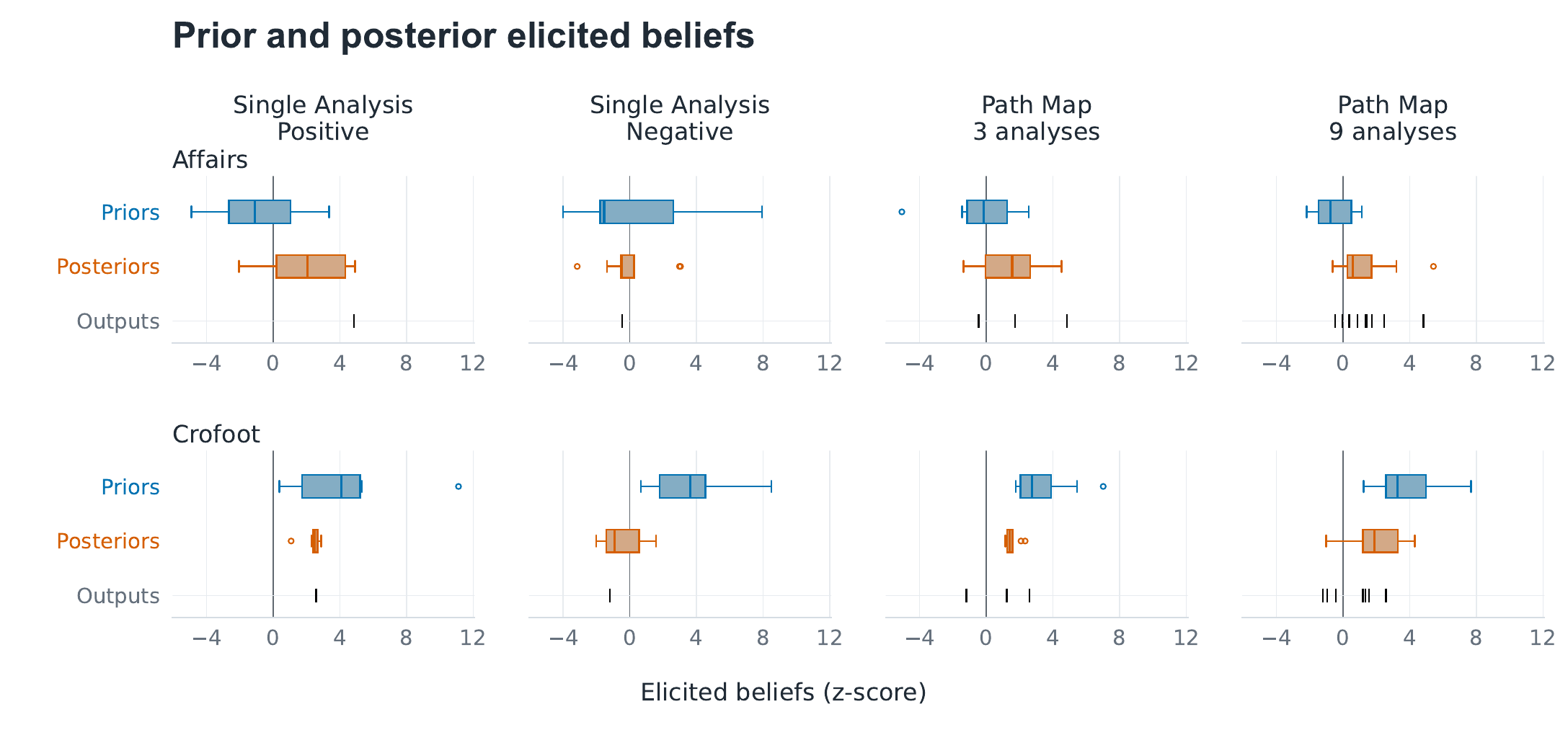}
  \caption{Participant belief shifts. Participants' posterior beliefs (orange) shift from their prior beliefs (blue) differently based on which displayed value(s) they review (Outputs lines). When shown one output, the posterior shifts closer to the displayed extreme value. When shown multiple analysis paths, the posteriors place a lower Weight of Advice on the extremes and lie closer to the median value.}
  \Description{Eight panels compare distributions of participant belief estimates before and after reviewing analysis results. Rows represent the Affairs and Crofoot tasks. Columns represent the positive-output baseline, negative-output baseline, three-analysis treatment, and nine-analysis treatment. Each panel contains an upper box plot of initial estimates, a lower box plot of final estimates, as well as reference tick marks beneath them indicating the displayed analysis results. The shared horizontal axis shows belief estimates as z-scores, with labeled ticks from minus 4 to 12 and a vertical line at zero. Boxes span the middle half of estimates, with an internal median line. Whiskers extend to the most extreme observations within one and a half interquartile ranges of the box. Isolated points show the observations past the whiskers. Final estimates are more tightly clustered in most panels, especially the Crofoot positive-output baseline and three-analysis treatment.}
 \label{fig:woa_data}
\end{figure*}

\begin{figure*}
  \centering
  \includegraphics[width=\linewidth]{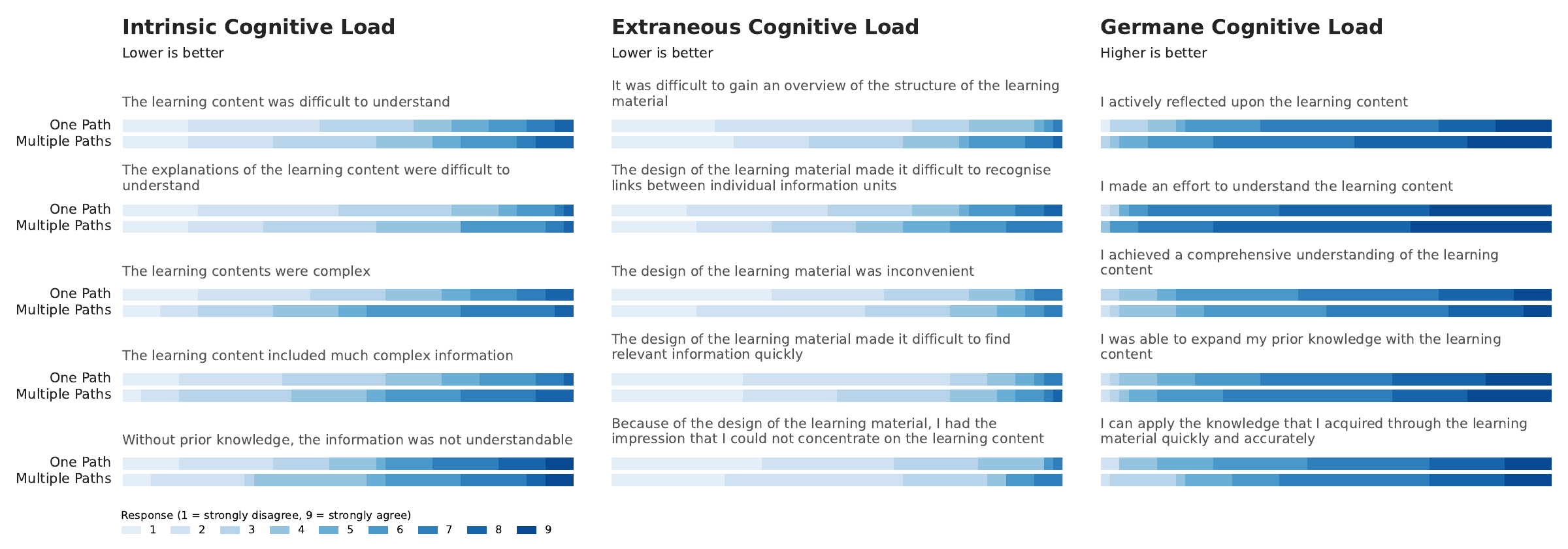}
  \caption{Cognitive Load questionnaire~\cite{krieglstein2023development}. After viewing alternative paths, participants reported higher Intrinsic and Extraneous Cognitive Load. They also reported higher Germane Cognitive Load, a beneficial cognitive engagement associated with learning. 
  }
  \Description{Stacked bar charts of cognitive load questionnaire responses. Three columns show Intrinsic, Extraneous, and Germane Cognitive Load, with five items per column. Each item has an upper bar for One Path and a lower bar for Multiple Paths. Their segment widths show response counts on a common horizontal scale. Ratings run from 1, not at all applicable, to 9, fully applicable, ordered from left to right. These ratings are shaded progressively darker. Lower ratings are labeled better for intrinsic and extraneous load; higher ratings are better for germane load.}
 \label{fig:likert}
\end{figure*}

Detailed results for the preregistered hypotheses comparing viewing one or multiple analyses can be found in Table~\ref{tab:main_results} and Figure~\ref{fig:posterior_draws}.
We additionally visualize participants' updated beliefs (Figure~\ref{fig:woa_data}), Cognitive Load questionnaire responses (Figure~\ref{fig:likert}), and preference rating responses (Figure~\ref{fig:preference_hist}).

\vspace{0.5em}
\noindent\textbf{[H1]}
We find a 99.92\% probability of support for H1, with a very large effect size of -0.63 regarding participants placing a lower Weight of Advice on extreme outputs when shown alternatives.

\vspace{0.5em}
\noindent\textbf{[H2]}
We find a 96.45\% probability of support for H2, with a large effect size of -0.38 regarding participants' final beliefs ending closer to the multiverse median when shown alternatives.

\vspace{0.5em}
\noindent\textbf{[H3]}
We find a 73.85\% probability of support for H3, with a very small effect size of 0.09 regarding participants having a wider confidence interval when shown alternatives.

\vspace{0.5em}
\noindent\textbf{[H4]}
We find a 100\% probability of support for H4, with a very large effect size of 0.96 regarding participants spending longer on the interface when shown alternatives.

\vspace{0.5em}
\noindent\textbf{[H5]}
We find a 99.35\% probability of support for H5, with a very large effect size of 0.40 regarding participants reporting a higher Intrinsic Cognitive Load when shown alternatives.

\vspace{0.5em}
\noindent\textbf{[H6]}
We find a 95.22\% probability of support for H6, with a large effect size of 0.31 regarding participants reporting a higher Extraneous Cognitive Load when shown alternatives.

\vspace{0.5em}
\noindent\textbf{[H7]}
We find an 82.38\% probability of support for H7, with a small effect size of 0.17 regarding participants reporting a higher Germane Cognitive Load when shown alternatives.

\subsection{Sub-condition analyses}
\label{sec:results-expl} 
Additional analyses compare between subconditions, for which detailed results can be found in Table~\ref{tab:expl_results} and Figure~\ref{fig:posterior_draws}.

\subsubsection{Seeing more alternatives shows diminishing returns.} 

Most trends remained the same, but with a lower magnitude, as in the main hypotheses when comparing between viewing three or nine analyses. There remained medium effects on Weight of Advice (-0.20), closeness of the posterior to the multiverse median (-0.24), and increased Intrinsic Cognitive Load (0.22). Unlike the main hypotheses, we find a small effect that seeing more alternatives reduced reported Germane Cognitive Load (-0.14).
Results suggest that \textit{samples} could provide a middle ground when optimizing for both understanding and effort.

\subsubsection{Overreliance and cognitive engagement reduce when the result disagrees with the prior.}

We find large effects that participants reported a higher Intrinsic Cognitive Load (0.38) and a lower Germane Cognitive Load (-0.33) when viewing a disagreeing output than an agreeing one. We also find a medium effect that participants placed a lower WoA when viewing a disagreeing output (-0.22). 
Results suggest that users do not engage uniformly with individual results.

\subsection{Qualitative results}
\label{sec:results-qual} 
From participants' reflections, we find they follow varied sensemaking strategies, rarely considered alternatives on their own, and wished for additional \textit{explanations} and \textit{interactions}.

\subsubsection{Participants followed a variety of self-described strategies to make sense of multiple outcomes.} 

Participants interpreted multiple outputs by widening uncertainty [P25, P26, P29, P30, P33, P34, P36], anchoring to the median analysis [P24, P30, P32, P39, P41], trying to find the single ``correct'' analysis [P23, P33, P42, P43], discounting the analyses [P11], and choosing the analysis that most reflected their beliefs [P42]. P3 and P28 raised concerns about $p$-hacking.

Some participants felt more agency seeing alternatives [P23, P26, P27, P37, P43]. For instance, P37 shared that \textit{``I felt like I did more work while I was working on [the treatment interface], instead of ... relying on the AI that this is the only way to do it.''} P23 liked that they could \textit{``critically think myself of why a method was chosen,''} and P26 appreciated that without a \textit{``definitive answer to the analysis question [I have to] like, pause and think about my decisions.''}

\subsubsection{Participants rarely thought of specific alternative analysis pathways on their own.}

Current practice expects users to notice, question, and provide alternatives for agent decisions. However,  participants rarely suggested alternative methods of analysis construction when asked what they would attempt next. Only six suggested an alternative covariate [P10, P17, P29, P33] or model [P6, P19] specifically, with ten wanting to adjust covariates [P4, P34, P48] or models [P12, P14, P18, P22, P34, P36, P38, P47] generically. 
P47 wanted to test different models, but found thinking of alternatives on their own challenging: \textit{``I think I would maybe try different analysis approaches? I'm not sure what exactly those would entail because I'm kind of rusty on my pulling out suitable analysis methods off the top of my head.''}

\subsubsection{Participants wished for additional explanations and interactions.} 

\begin{figure}
  \centering
  \includegraphics[width=\linewidth]{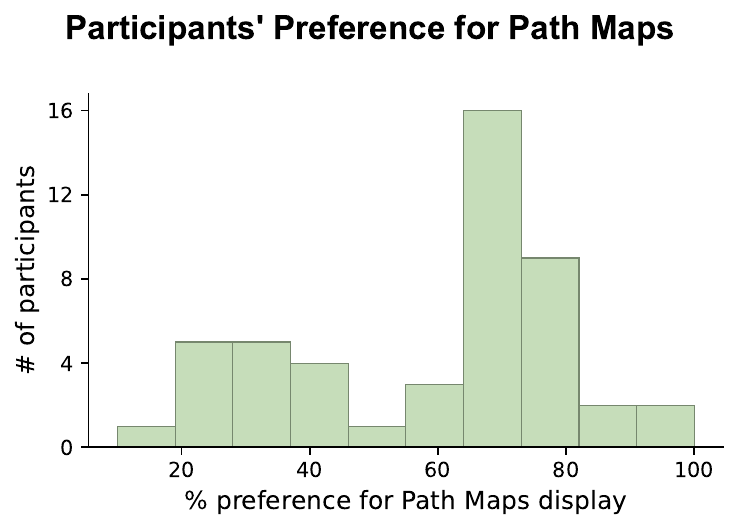}
  \caption{Participant preference for the Path Maps display. Participants rated each interface by preference such that the ratings added to 100. Some participants highly preferred each display, with the majority preferring the Path Maps display.}
  
  \Description{A histogram of participants' preference for the Path Maps interface. The horizontal axis gives the percentage of each participant's preference allocated to the Path Maps interface, with the remaining percentage being allocated to the single-analysis interface.  An allocation of 50 percent indicates equal preference. The vertical axis gives the number of participants, with label ticks ranging from 0 to 16. The graph comprises ten adjacent bins. The tallest bar, near 70 percent, contains sixteen participants. The neighboring bar to its right contains nine. A smaller concentration lies between about 20 and 40 percent, and few participants are near either endpoint.}
 \label{fig:preference_hist}
\end{figure}

While two-thirds of participants preferred the treatment interface (Figure~\ref{fig:preference_hist}), they suggested improvements for \textbf{explanations} and \textbf{interactions}.
\textit{Navigational} supports for filtering paths [P4, P40] and for toggling individual alternatives [P1, P3, P34, P38] would aid comparison, particularly alongside additional \textit{difference highlighting}
[P4, P7, P22, P36, P40].
\textit{Details on demand} \textbf{interactions} would enable an initial simpler overview [P5, P8, P9, P12, P20].
Some participants suggested more descriptive [P6, P22, P26, P34] and opinionated [P6, P37] rationales, including the agent's \textit{preferred choice} [P24, P43]. P43 would \textit{``like the AI to say which it recommended out of those combinations so that I'd be able to see which one it wanted most, but also, in case I disagree with it, I can see the different combinations as well.''}

\section{Discussion}
\label{sec:discussion}
Agentic Path Maps improve understanding of diverging outputs on underspecified, automatically executed, agentic tasks. 
We discuss tasks with a higher utility for Agentic Path Maps (\ref{sec:dis-when}), methods for evaluating Agentic Path Maps (\ref{sec:dis-eval}), and promising directions for studying the design space (\ref{sec:dis-ds}).

\subsection{Weighing the benefits of additional choice against higher costs}
\label{sec:dis-when}

Agentic Path Maps provide both benefits (e.g., surfacing alternatives and less overreliance) and drawbacks (e.g., higher computational cost and Intrinsic Cognitive Load). The benefits of Agentic Path Maps are particularly salient when tasks are high-stakes, responsive to user context, and involve many criteria, as they enable users to retain agency over the final output while reducing interruptions. 
When privacy is a concern, Agentic Path Maps enable personalization without modeling the user. 
However, other approaches serve better when users want to be iteratively engaged on a bespoke platform, contradictions to user context are low-stakes, or there is one task-valid output. 

The Paradox of Choice~\cite{schwartz2004paradox} highlights a tension regarding the ideal number of choices in psychology. Without choices, people may not have their needs met, but excessive choices cause fatigue and reduce subjective experience.
``Choosing when to choose''~\cite{schwartz2004paradox}, by making high-level decisions about what low-level decisions to attend to, can reduce this tension and is reflected in both psychology~\cite{huys2012bonsai,schwartz2004paradox} and hierarchical planning~\cite{russell1995artificial}. 
Navigating this paradox is a rich path for future work, as optimal approaches will likely vary among domains and users.

\subsection{Evaluating agentic systems for unique users who are not static oracles}
\label{sec:dis-eval}

We need methods to evaluate if agents can create Agentic Path Maps and if human reviewers can interpret them. 
Rather than using a dichotomous definition of correctness, evaluation of Agent-created Path Maps must review coverage of the task-valid set. BLADE's coverage@k metric is a rare example of this breadth-focused approach~\cite{gu2024blade}. 
Unlike standard approaches to interpretability research that define a correct and incorrect answer to measure appropriate reliance~\cite{vasconcelos2023explanations,bansal2021does,buccinca2021trust}, evaluating human interpretation of Agentic Path Maps must consider correctness as unique to individual users.
Rather than considering humans as oracles who can be queried at an optimal rate~\cite{majumder2024data,dong2026value,chan2017semantically,kraus2020effects}, we must recognize that users may not know their preferences, want to explore multiple options, or change their underlying preferences when asked explicitly~\cite{schwartz2004paradox,grunde2026overseeing}.

\subsection{Furthering recommendations in the design space}
\label{sec:dis-ds}

We encourage refinement through empirical study of recommendations for design space dimensions. Our user study results encourage further development of \textbf{explanatory content}, \textbf{interactions}, and what \textit{sample} to display. 
In open-ended environments, agents' decision spaces can explode in size, motivating study of who takes \textbf{initiative} for determining the \textit{granularity} and \textit{quantity} of content. 
Finally, engaging with the broader design space of displaying multiple items beyond Agentic Path Maps could build fruitful systems with mixed approaches to human involvement.
 
\section{Limitations}
\label{sec:limitations}
Our work contains several limitations. Our user study uses convenience sampling from an academic institution for a participant sample size within our resource constraints. 
We study human-made decision paths weighted equally. Future study of Agentic Path Maps made by agents will complicate the determination of whether paths are task-valid. 
We study the data analysis domain, limiting generalizability to other domains. Data analysis stimuli were challenging to create, and some participants expressed confusion and reported unlikely belief elicitation values. 
Additionally, we did not concretely measure the resulting increase in computational costs. The extra cost of creating alternatives should be further understood, particularly for irreversible or purchasing tasks.
Finally, there remain ethical considerations in evaluating what a ``complete'' task-valid set entails and how to incorporate that definition into evaluation.

\section{Conclusion}
With Agentic Path Maps, humans can review a set of task-valid alternatives, revealing their user-valid outputs to under-specified tasks without excessive interruptions. Our design space enables designers to implement and study this approach, and our empirical results demonstrate how users can be misled when viewing one output, even if that output is correct under the task constraints. In this work, we encourage and enable future study on designing Agentic Path Maps to improve human control of agentic outputs.

\begin{acks}
This research was supported in part by NSF award IIS-2402647 and a gift from Google. Thank you to Brandon Kill and William Kumler for feedback on statistical testing procedures. 
\end{acks}

\bibliographystyle{ACM-Reference-Format}
\bibliography{0_main}
\end{document}